\documentclass[aps,prl,twocolumn,superscriptaddress,citeautoscript,amssymb,reprint,showpacs,floatfix,longbibliography]{revtex4-2}

\usepackage{graphicx}% Include figure files
\usepackage{dcolumn}% Align table columns on decimal point
\usepackage{amsmath}
\usepackage[version=4]{mhchem}
\usepackage{xcolor}
\usepackage{hyperref}
\usepackage[T1]{fontenc}

\begin{document}

\preprint{APS/123-QED}

\title{Concurrence of directional Kondo transport and incommensurate magnetic order in the layered material \ce{AgCrSe2}}%

\author{José Guimarães}
\affiliation{Max Planck Institute for Chemical Physics of Solids, 01187 Dresden, Germany}
\affiliation{School of Physics and Astronomy, University of St Andrews, St Andrews KY16 9SS, UK}
\author{Dorsa S. Fartab}
\affiliation{Max Planck Institute for Chemical Physics of Solids, 01187 Dresden, Germany}

\author{Michal Moravec}
\affiliation{Max Planck Institute for Chemical Physics of Solids, 01187 Dresden, Germany}
\affiliation{School of Physics and Astronomy, University of St Andrews, St Andrews KY16 9SS, UK}

\author{Marcus Schmidt}
\affiliation{Max Planck Institute for Chemical Physics of Solids, 01187 Dresden, Germany}

\author{Michael Baenitz}
\affiliation{Max Planck Institute for Chemical Physics of Solids, 01187 Dresden, Germany}

\author{Burkhard Schmidt}
\affiliation{Max Planck Institute for Chemical Physics of Solids, 01187 Dresden, Germany}

\author{Haijing Zhang}
\email{Correspondence: haijing.zhang@cpfs.mpg.de}
\affiliation{Max Planck Institute for Chemical Physics of Solids, 01187 Dresden, Germany}

%\date{\today}

\begin{abstract}

\section{Abstract}

In this work, we report on the concurrent emergence of the directional Kondo behavior and incommensurate magnetic ordering in a layered material. We employ temperature- and magnetic field-dependent resistivity measurements, susceptibility measurements, and high resolution wavelength X-ray diffraction spectroscopy to study the electronic properties of \ce{AgCrSe2}. Impurity Kondo behavior with a characteristic temperature of $T_\text K=32\,\rm K$ is identified through quantitative analysis of the in-plane resistivity, substantiated by magneto-transport measurements. The agreement between our experimental data and the Schlottmann’s scaling theory allows us to determine the impurity spin as $S = 3/2$. Furthermore, we discuss the origin of the Kondo behavior and its relation to the material's antiferromagnetic transition. Our study uncovers an unusual phenomenon---the equivalence of the Néel temperature and the Kondo temperature---paving the way for further investigations into the intricate interplay between impurity physics and magnetic phenomena in quantum materials, with potential applications in advanced electronic and magnetic devices.

\end{abstract}

\maketitle

\section{Introduction}

Layered quantum materials, consisting of alternating sheets of magnetic rare earth or transition metal ions on a triangular lattice and nonmagnetic transition metal planes, comprise a unique combination of magnetic frustration and highly anisotropic transport properties. These materials exhibit a plethora of unconventional properties, such as various types of magnetic order~\cite{damay:13,matsuda:20}, giant magnetoresistance~\cite{PhysRevLett.61.2472,Ideue2020} and skyrmions~\cite{fert2017magnetic}, which arise from the combined effects of the magnetic order and electron correlations. Investigating how the electronic degrees of freedom couple with the magnetic order in such systems is key to understanding the microscopic mechanisms of the observed physical phenomena.

\ce{AgCrSe2} is a member of the class of intercalated transition metal dichalcogenides where \ce{CrSe2} transition metal dichalcogenide layers alternate with transition metal Ag layers. It has been recently demonstrated to be a promising thermoelectric material~\cite{Wu2016,Shiomi2018,hua2021}, a spin-orbit derived giant magnetoresistance~\cite{takahashi2022} and a spontaneous anomalous Hall effect driven by topological effects~\cite{kim2023} have been reported in this system as well. The Cr atoms host $S=3/2$ spins arranged on a triangular lattice in each layer. The competing exchange interactions between the first- and third-nearest neighbour Cr moments lead to an incommensurate magnetic order at $T_\text N\approx32\,\rm K$~\cite{baenitz2021}.

In this article, we report the concurrent emergence of Kondo behaviour in the temperature and field dependence of the in-plane electrical resistivity and incommensurate magnetic order of the Cr spins in \ce{AgCrSe2}. On the same temperature scale as this magnetic ordering, impurity Kondo behaviour emerges. Essential to understand this is the $\approx1\,\%$ off-stoichiometry of the Cr ions, detected by high-resolution wavelength X-ray diffraction spectroscopy, in our sample. The hybridization of the charge carriers at the Fermi level, dominated by the Se 4$p$ states, with the Cr 3$d^3$ states of these excess impurities would in principle lead to Kondo physics at low enough temperatures. However, above $T_\text N$, strong in-plane ferromagnetic fluctuations inhibit the hybridization between itinerant electrons and excess impurities. Only in the incommensurate magnetic order regime, at lower temperatures, an effective Kondo hybridization can exist. The key finding of our work--- Kondo screening is enabled and associated with the antiferromagnetic order--- opens an alternative route to tune the competition between the magnetic correlations and Kondo effect, which plays a central role in the heavy-fermion systems~\cite{hewson1993} and unconventional superconductors~\cite{Nagaosa1997,RevModPhys.84.1383,paschen2021quantum}.

\begin{figure*}[t!]
\centering
    \includegraphics[width=18cm]{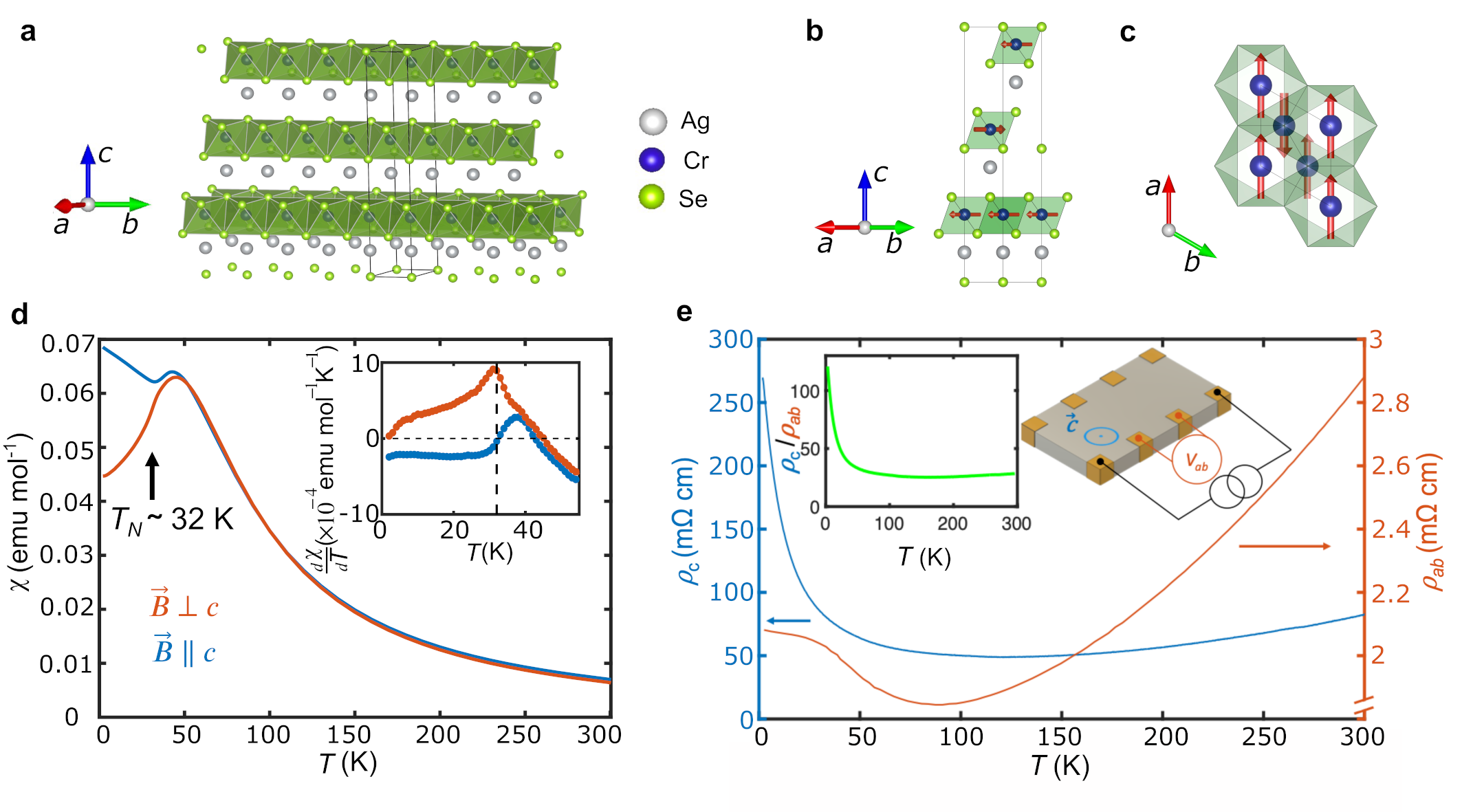}
    \caption{\textbf{2D crystal structure and characteristic resistivity of \ce{AgCrSe2}.} \textbf{a} Crystal structure of \ce{AgCrSe2}, space group R3m. The translucent green polyhedra illustrate the tilted and distorted edge-sharing octahedra forming the triangular lattices. \textbf{b} Side view of the unit cell, including the \ce{Cr} atoms spin that arrange antiferromagnetically along the $c$ direction. \textbf{c} Bottom view of the unit cell showing the ferromagnetically alignment in the $ab$ plane, in this simplification the long wavelength cycloidal periodicity discussed in Baenitz et al.~\cite{baenitz2021} is ignored. \textbf{d} Magnetic susceptibility measured with a 1 T magnetic field in the $ab$ plane, orange line, and along the $c$ axis, blue line, and its derivative with respect to temperature in the inset, following the same color code. The black arrow is pointing to a kink in the magnetic susceptibility, revealing the Néel transition temperature, $T_\text N$. This is the same temperature represented by the vertical black line depicted in the inset. \textbf{e} Temperature dependence of the resistivity of the \ce{AgCrSe2} single crystal in the $ab$ plane, orange line, along the $c$ axis, blue line, and their ratio in the inset, green line.}
    \label{fig1}
\end{figure*}

\section {Results}

\subsection*{Electronic and magnetic structure of \ce{AgCrSe2}}

High quality \ce{AgCrSe2} single crystals were obtained via chemical transport reaction, and the growth process is described in detail in Baenitz et al.~\cite{baenitz2021}. The crystal structure of the \ce{AgCrSe2} single crystal is comprised of alternating layers of \ce{Ag} and edge-sharing tilted distorted \ce{CeSe6} octahedra~\cite{li2018,wang2020,baenitz2021,gesa2023} that repeat along the crystallographic \textit{c} axis (Fig.~\ref{fig1}a-c). This crystal structure is very similar to the one that characterizes delafossite metals (R$\overline{3}$m)~\cite{Mackenzie2017}, however, there is no centro-symmetric inversion symmetry in this space group: R3m. The Cr atoms host $S=3/2$ spins arranged on a triangular lattice in each layer~\cite{baenitz2021}.  Neutron measurements evidence an incommensurate magnetic order below $T_\text N = 32\,\rm K$ with ordering vector ${\bf Q}=(0.037,0.037,3/2)$ in units of the reciprocal lattice constants. Below $T_\text N$, the Cr$^{3+}$ moments order almost ferromagnetically in the triangular plane (positive single-ion anisotropy) in a cycloidal fashion with an antiferromagnetic stacking perpendicular to the plane~\cite{baenitz2021}. This value for the Néel temperature~\cite{Fisher1962} is consistent with the temperature dependence of the magnetic susceptibility, $\chi$, depicted in Fig.~\ref{fig1}d, and previously discussed in \cite{baenitz2021,kim2023}.
High quality \ce{AgCrSe2} single crystals were obtained via chemical transport reaction, and the growth process is described in detail in Baenitz et al.~\cite{baenitz2021}. The crystal structure of the \ce{AgCrSe2} single crystal is comprised of alternating layers of \ce{Ag} and edge-sharing tilted distorted \ce{CeSe6} octahedra~\cite{li2018,wang2020,baenitz2021,gesa2023} that repeat along the crystallographic \textit{c} axis (Fig.~\ref{fig1}a-c). This crystal structure is very similar to the one that characterizes delafossite metals (R$\overline{3}$m)~\cite{Mackenzie2017}, however, there is no centro-symmetric inversion symmetry in this space group: R3m. The Cr atoms host $S=3/2$ spins arranged on a triangular lattice in each layer~\cite{baenitz2021}.  Neutron measurements evidence an incommensurate magnetic order below $T_\text N = 32\,\rm K$ with ordering vector ${\bf Q}=(0.037,0.037,3/2)$ in units of the reciprocal lattice constants. Below $T_\text N$, the Cr$^{3+}$ moments order almost ferromagnetically in the triangular plane (positive single-ion anisotropy) in a cycloidal fashion with an antiferromagnetic stacking perpendicular to the plane~\cite{baenitz2021}, as depicted in Fig.~\ref{fig1}b,c. This value for the Néel temperature~\cite{Fisher1962} is consistent with the temperature dependence of the magnetic susceptibility, $\chi$, depicted in Fig.~\ref{fig1}d, and previously discussed in Baenitz et al. and Kim et al.~\cite{baenitz2021,kim2023}.

\subsection*{Anisotropic transport properties of \ce{AgCrSe2} under the influence of magnetic field}

In order to study the electrical transport properties of the single crystal, a standard four-probe method was used in measuring a millimeter sized sample~\cite{vdp,zhang2019band}. In Fig.~\ref{fig1}e, we present the temperature dependence of the resistivity both in the \textit{ab} plane and perpendicular to it (note the different scales). We observe a strong anisotropy in the resistivity (inset of Fig.~\ref{fig1}e), along the \textit{c} direction, at $T\approx2\,\rm K$, the resistivity is around two orders of magnitude higher than the in-plane resistivity, as expected for such a quasi two-dimensional (2D) layered compound  ~\cite{valla2002}. This anisotropy has been previously discussed~\cite{baenitz2021,yano2016} and suggests that the origin of the current along \textit{c}, at the temperature range considered, are thermally activated Se states that dominate the density of states near the Fermi surface~\cite{baenitz2021,hua2021}.

The resistivity along the \textit{c} axis shows a typical semiconductor behaviour: metallic-like temperature dependence at higher temperatures with the resistivity decreasing with decreasing temperature up to the point, at around $80\,\rm K$, where the thermally activated conductors freeze out, revealing the intrinsic insulating character of the material. On the other hand, the \textit{ab} plane resistivity depicts an interesting and yet to be discussed feature: Similar to the \textit{c} plane resistivity, the \textit{ab} plane resistivity decreases with the decrease of temperature and at lower temperatures starts increasing. This would again be typical of a semiconductor, however here we observe that this low-temperature increase saturates, motivating us to perform a more detailed investigation of the in-plane resistivity at lower temperatures. 

\begin{figure*}[t!]
\centering
    \includegraphics[width=18cm]{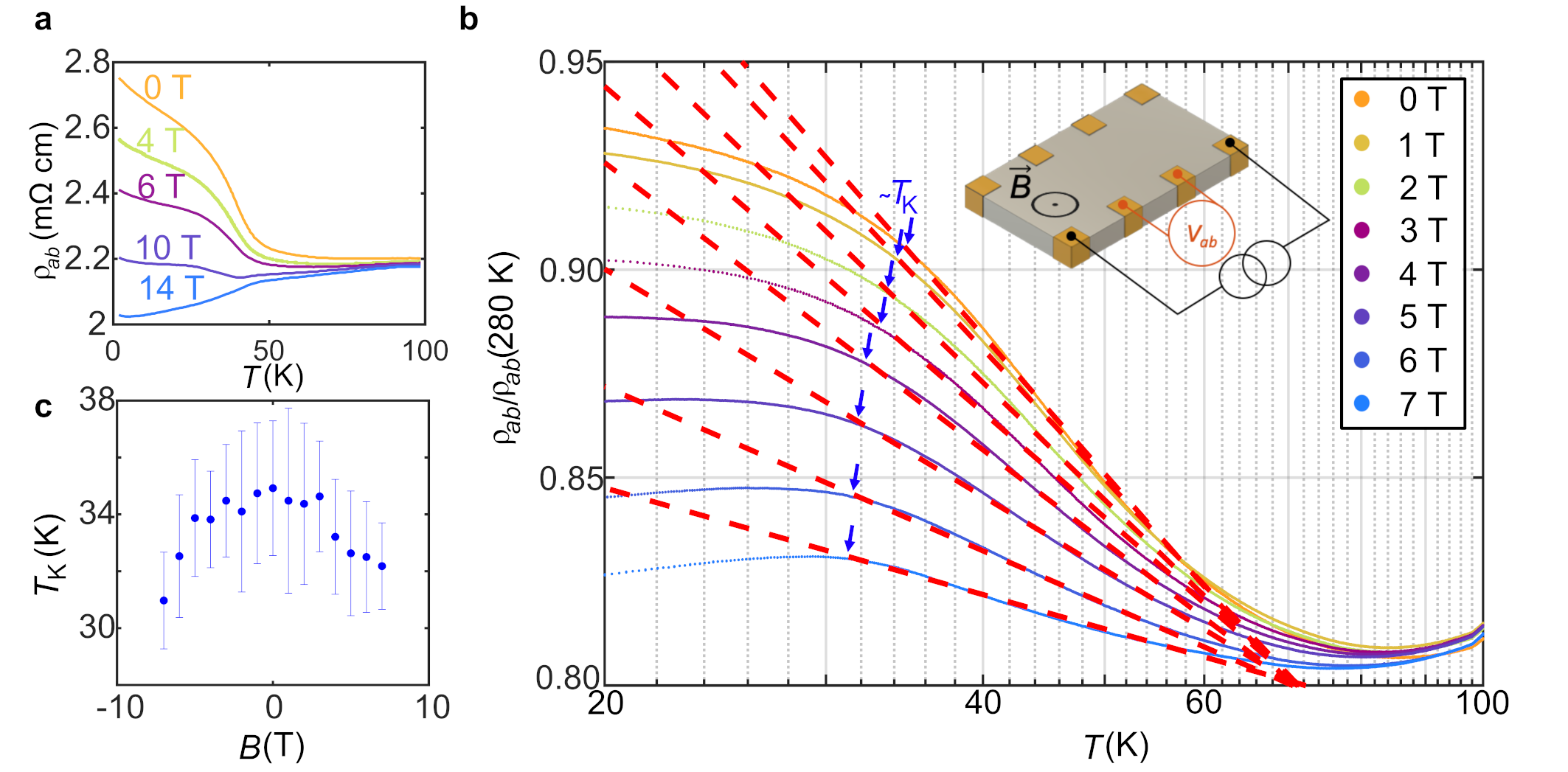}
    \caption{\textbf{In-plane resistivity of \ce{AgCrSe2} as a function of temperature under perpendicular magnetic fields.} \textbf{a} $ab$ plane resistivity as a function of temperature with different applied out of plane magnetic fields $\mu_0H=0\dots14\,\rm T$. With a large enough field of $\mu_0H\gtrsim14\,\rm T$ metallic behaviour can be observed in the entire temperature range $2\,{\rm K}\le T\le100\,\rm K$. \textbf{b} Normalized $ab$ plane resistivity as a function of temperature, showing the $-\ln(T)$ dependence at intermediate temperatures characteristic of the Kondo effect. The red dashed lines correspond to this logarithmic temperature dependence. The blue arrows point at the approximate value of  $T_\text K$, the Kondo temperature value is estimated as the temperature at which the resistivity deviates from the logarithmic behaviour. \textbf{c} Variation of the Kondo temperature estimated from the resistivity measurements as a function of the out of plane magnetic field $\mu_0H=-7\,{\rm T}\dots7\,\rm T$, showing a small and symmetric variation around $H=0$. The values shown are an average of repeated measurements and the error bars represent the standard error of the mean.}
    \label{fig2}
\end{figure*}

We now focus on the resistivity increase and saturation at temperatures below the resistivity minimum. The \textit{ab} plane resistivity measurements as a function of temperature were repeated at different values of an applied magnetic field perpendicular to the triangular plane. From Fig.~\ref{fig2}a,b, we see that by applying a magnetic field the upturn in resistivity with the decrease of temperature is gradually suppressed. Furthermore, for high enough fields, 14 T, the upturn is fully suppressed with \ce{AgCrSe2} showing a metallic behaviour in the entire temperature range. The possibility of manipulating the phenomena with field indicates that the origin of the upturn might be magnetic in nature. 

\subsection*{Single-ion Kondo effect in the \textit{ab} plane resisitivity of \ce{AgCrSe2}}

A possible mechanism to explain the data observed is the single-ion Kondo effect~\cite{kondo1964}. As an experimentally motivated model, it was first used to describe the observation that the resistivity of some metals containing magnetic impurities deviates at low temperatures from a monotonous decrease but instead has a minimum at a certain temperature. Furthermore, at temperatures below the minimum a logarithmic increase is observed followed by saturation. This increase can be attributed to the appearance of additional magnetic exchange scattering between electrons at the Fermi level and localized magnetic moments substantially below it. Computing the scattering probability for conduction electrons in the Kondo model, a $-\ln(T)$ increase in resistivity with temperature is obtained~\cite{kondo1964} below the minimum temperature. Furthermore, as temperature approaches $T\gtrsim0\,\rm K$, there is a crossover from the $-\ln(T)$ scaling~\cite{abrikosov1965} towards saturation.

In order to discuss the viability of the single-ion Kondo model to describe the resistivity behaviour observed, the resistivity was measured at magnetic fields between $-9\,\rm T$ and $9\,\rm T$, observing that for fields higher than $7\,\rm T$, there is no visible Kondo effect. For fields between $-7\,\rm T$ and $7\,\rm T$, there clearly is a $-\ln(T)$ dependence for temperatures below the temperature at which the resistivity is minimum.

The resistivity for very low temperatures deviates from the $-\ln(T)$ scaling: this crossover to saturation can be taken as an approximation of the Kondo temperature $T_\text K$~\cite{suhl1965,zhu2016,khadka2020}. With the rich variety of systems revealing Kondo-like transport, there have been a wide number of numerical and analytical techniques that aim to quantitatively determine the Kondo temperature based on resistivity measurements~\cite{suhl1965,tsvelick1983,coleman1984,bickers1987}. Innocently looking, obtaining quantitative results from the Kondo model is a seemingly complex task and often requires bold simplifications. In this sense, the determination of the Kondo temperature scale from only the resistivity behaviour should be taken as an approximate estimate rather than an exact temperature.

From the electrical transport measurements, the Kondo temperature of \ce{AgCrSe2} is $T_\text K\approx34\,\rm K$ at $\mu_0H=0$ T. Although the magnetic field suppresses the upturn, the variation of the Kondo temperature itself with different magnetic fields is very small, see Fig.~\ref{fig2}c, and shows that the strength of the Kondo effect remains essentially robust with field until it is fully suppressed at a certain critical field. The resistivity measurements were performed in magnetic fields in both perpendicular directions to the \textit{ab} plane to exclude inhomogeneity and misalignment effects. 

The behaviour of the resistivity in applied magnetic fields shown in Fig.~\ref{fig2}b is also qualitatively explained ~\cite{otte2008} in terms of the Kondo effect, which is suppressed when the external applied magnetic field polarizes the impurity's spin, i.\,e., when the Zeeman energy of the impurity spin is of the order of the characteristic energy scale for the Kondo effect~\cite{zitko2009}. In our case, with an impurity spin $S=3/2$ and a characteristic temperature $T\textsubscript{K}\approx34\,\rm K$, the corresponding crossover field necessary to suppress the Kondo effect is $\mu_0H\approx8\,\rm T$, which is in good agreement with the observed results.

\subsection*{Magnetic field dependence of the Kondo effect: Schlottmann's scaling}

The Kondo effect can also be observed in the magnetic field dependence of the resistivity of \ce{AgCrSe2}. According to Schlottmann’s work on the Kondo model~\cite{schlottmann1983}, the exchange scattering is dominated by a single energy scale related to $k_\text B T_\text K$ if the Kondo behaviour originates from scattering off localised magnetic impurities. This relation provides a much better definition for $T_\text K$ and it is a property unique to the impurity Kondo model. By solving the Coqblin-Schrieffer Hamiltonian~\cite{coqblin1969} using the Bethe Ansatz technique, for a given value of the impurity spin $S$ the magnetoresistivity of a Kondo system was shown to have a universal scaling behaviour with field and temperature. The scaling field $B^{\ast}(T)$ depends on the Kondo temperature $T_\text K$ according to $B^{\ast}(T)=B^{\ast}(0)+k_\text BT/(g\mu)=k_\text B(T_\text K+T)/(g\mu)$ where $g$ is the Landé factor and $\mu$ is the expected moment of the localised magnetic impurity.

\begin{figure*}[t!]
\centering
    \includegraphics[width=18cm]{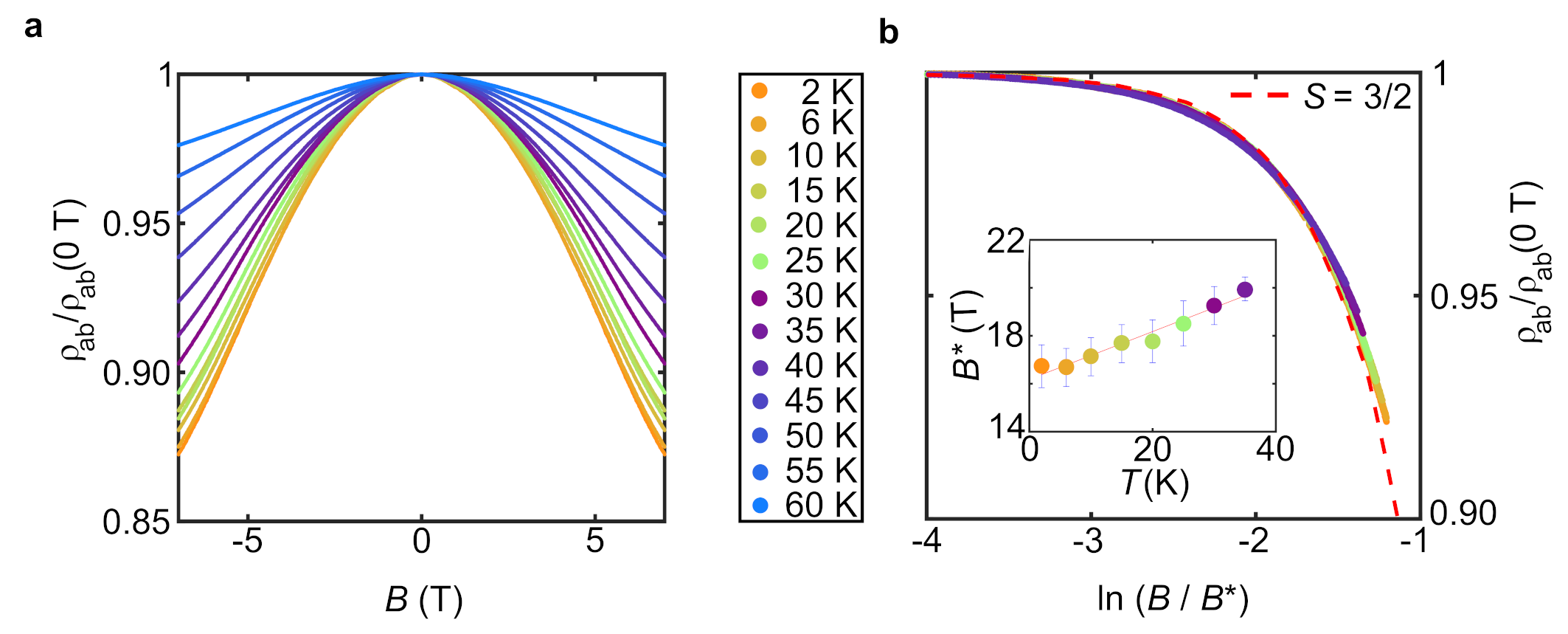}
    \caption{\textbf{Schlottmann's scaling and magnetoresistivity.} \textbf{a} Normalized resistivity as a function of magnetic field at different temperatures $T$ between $2\,{\rm K}$ and $60\,\rm K$. The resistivity has been measured along the $ab$ plane and normalised to the resistivity measured for each temperature with no applied magnetic field. \textbf{b} Normalized magnetoresistivity as a function of $\ln(B/B^*)$ for fields between $-5\,\rm T$ and $5\,\rm T$ and temperatures up to $35\,\rm K$. The fit parameter $B^*$ is the scaling field, discussed in the 'Magnetic field dependence of the Kondo effect: Schlottmann’s scaling' subsection of the Results. The theoretically expected curve~\cite{schlottmann1983} is marked by a red dashed line. Inset to \textbf{b} Scaling field $B^*(T)$ obtained from the fit of the normalized resistivities to the theoretical curve for temperatures from $2\,\rm K$ up to $35\,\rm K$. The error bars represent the values of the scaling field within which the coefficient of determination between the theoretical curve and the scaled experimental data is higher or equal than 0.95.}
    \label{fig3}
\end{figure*}

We make use of this universal relation as follows. First we measure the field dependence of the resistivity for different temperatures below the Kondo minimum (as depicted in Fig.~\ref{fig3}a) and scale them to the theoretical curve calculated by Schlottmann~\cite{schlottmann1983}. By fitting the scaled data with theoretical curves at different values of $S$, therefore $\mu$, the relation allows a determination of the impurity spin $S$ from the quality of the overlap between the scaled magnetoresistivity measured at each temperature and the theoretically expected value. The observed data scaling collapse is good, and agrees well with the scaling curve for $S=3/2$, shown in Fig.~\ref{fig3}b.

Second we obtain $B^{*}(0)$ from the scaling behaviour extrapolated to zero temperature. Knowing the value for $S$, that can be used to independently determine the Kondo temperature. The linear fit of the scaling field variation with temperature is shown in the inset of Fig.~\ref{fig3}b. In summary, using Schlottmann’s scaling, a value $S=3/2$ is consistent with the experimental data and a Kondo temperature $T_\text K=(32\pm2)\,\rm K$ is obtained.

\subsection*{Emergence of Kondo behaviour and the relation to the incommensurate magnetic order of the Cr spins}

In the section above, we have shown by measuring the temperature dependence of the electrical resistivity under a constant external magnetic field and the magnetoresistivity at different temperatures that the observed resistivity behaviour at low temperatures in the \textit{ab} plane shows the hallmarks of a Kondo system, with a characteristic temperature $T_\text K\approx32\,\rm K$. The Schlottmann scaling fit shown in Fig.~\ref{fig3}b indicates that the magnetic impurities that drive the effect have a spin $S=3/2$.

To investigate the origin of Kondo effect, wavelength X-ray dispersion measurements were performed in order to determine the crystal’s elemental composition and concentration with high accuracy. The results shown in Table~\ref{tab1} indicate that the crystal has an off stoichiometry of \ce{Ag_{0.942$\pm$0.004}Cr_{1.013$\pm$0.004}Se_2}. This composition was obtained taking into consideration the percentage weight at the perfect stoichiometry and normalized weights assuming a perfect concentration of Se. Furthermore the stoichiometry calculated is consistent with the carrier density obtained in Hall effect measurements~\cite{kim2023}.

The off-stoichiometry of Cr atoms is significant enough to be the origin of the Kondo effect, as only a very dilute concentration of the magnetic impurities is needed to observe a resistivity minimum~\cite{kondo2005}. Moreover, the Kondo effect has been previously reported elsewhere in antiferromagnetic materials with an excess of magnetic dopants~\cite{khadka2020}.

From Table~\ref{tab1} we learn that the observed excess of Cr ions is accompanied by a lack of Ag ions. Therefore the most natural locations for the extra Cr impurities are silver vacancies.

\begin{table}
\centering
    \caption{\textbf{Wavelength-dispersive X-ray measurement of single crystal \ce{AgCrSe2}}. The elemental information obtained from this spectroscopy measurement, weight and atomic percentage and their respective errors ($\Delta$), reveals that the crystal does not have a perfect stoichiometry: having extra Cr ions and lacking Ag ions.}
    \begin{tabular}{|c|c|c|c|c|}
        \hline
         & Weight (\%) & $\Delta$ Weight (\%) & Atomic (\%) & $\Delta$ Atomic (\%) \\
        \hline
        Ag & 33.043 & 0.117 & 24.815 & 0.063 \\
        \hline
        Cr & 17.132 & 0.044 & 25.616 & 0.054 \\
        \hline
        Se & 51.360 & 0.110 & 50.569 & 0.077 \\
        \hline
    \end{tabular}
    \label{tab1}
\end{table}

\begin{figure*} [t!]
\centering
    \includegraphics[width=18cm]{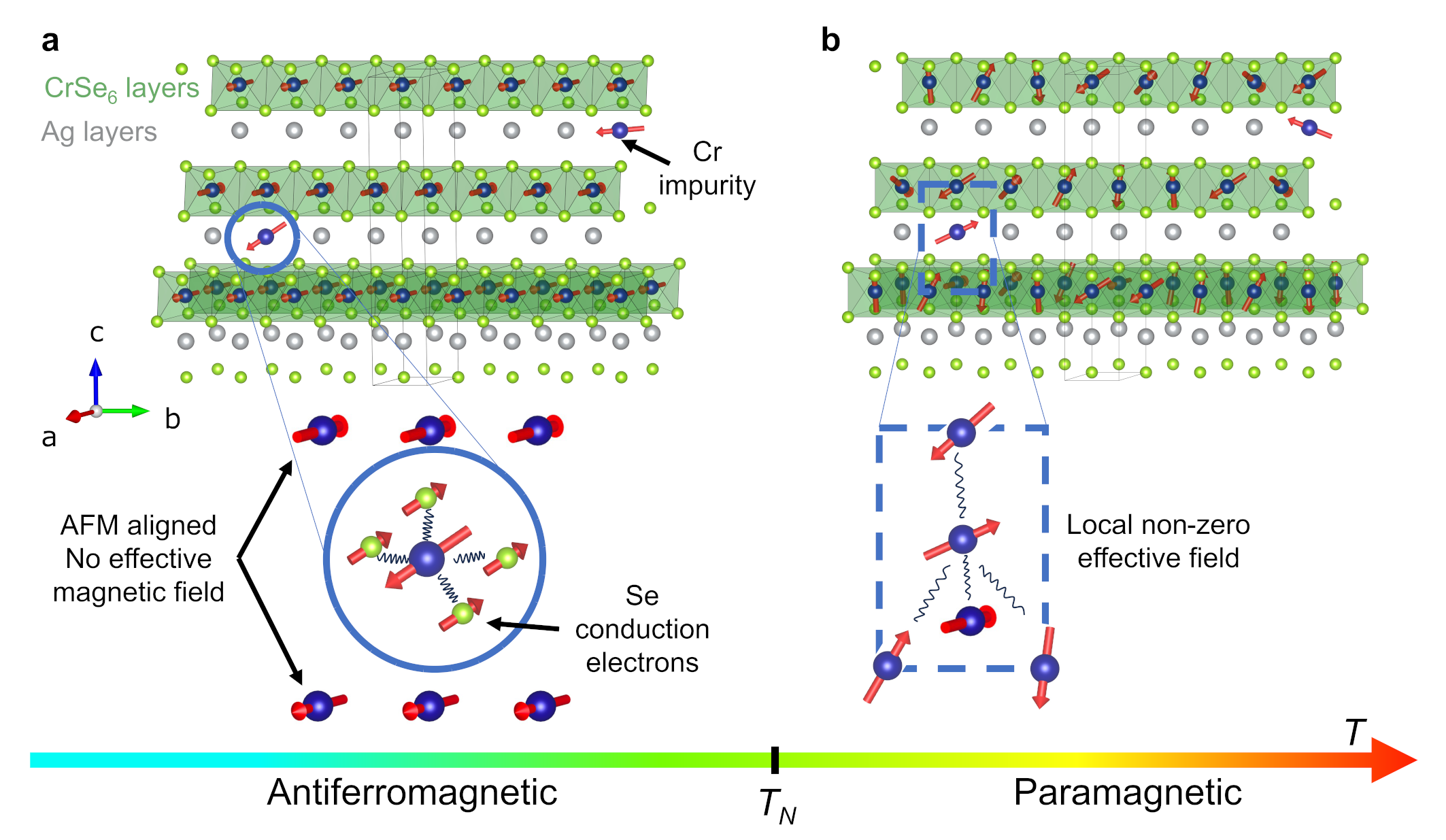}
    \caption{\textbf{Schematic of the magnetic interactions that determine the concurrence of Kondo effect and the incommensurate magnetic arrangement of the \ce{Cr} atoms.} Schematic illustration of the mechanisms behind the Kondo effect, highlighting the magnetic arrangements and interactions in play and how they differ in the \textbf{a} antiferromagnetic regime, at temperatures bellow the Néel temperature and \textbf{b} paramagnetic regime at higher temperatures. It is within this physical picture that the dependence of the Kondo regime on the antiferromagnetic transition becomes clear: as only at lower temperatures there is no effective magnetic field at the impurities sites and the impurity Kondo scattering can be observed. Simplification of the magnetic structure model, it does not reflect the cycloidal ordering discussed in Baenitz et al.~\cite{baenitz2021} which is not important in our context.}
    \label{fig4}
\end{figure*}

From the ARPES data~\cite{hua2021}, we know that the states near the Fermi surface are dominated by the Se $4p$ states. It is to those conduction bands that the impurity Cr atoms couple, leading to Kondo physics, as represented in the schematic in Fig.~\ref{fig4}.

\section{Discussion}

The puzzling question that we now tackle is why we have $T_\text N=T_\text K$ and why the magnetic order actually induces the Kondo physics we observe in our transport measurements. As far as we are aware, this is the first system where the Néel temperature and the Kondo temperature coincide. The Néel temperature is a critical temperature that determines a magnetic phase transition. In contrast, the Kondo temperature does not represent a phase transition but rather represents an energy scale for magnetic scattering of conduction electrons with magnetic impurities deep below the Fermi level, unrelated to magnetic order. 

The magnetic structure of stoichiometric \ce{AgCrSe2} is characterized by the Cr atoms that host a spin $S=3/2$ in a $3d^3$ configuration~\cite{baenitz2021,engelsman1973}. The Cr atoms in each layer form a triangular lattice and interact predominantly ferromagnetically within the \textit{ab} plane. Out of the plane, adjacent layers of Cr atoms couple antiferromagnetically. Magnetic susceptibility measurements~\cite{baenitz2021} yield an ordering temperature of $T_\text N=32\,\rm K$. 

In the paramagnetic phase above $T_\text N$, strong ferromagnetic fluctuations of the Cr spins in the \ce{CrSe2} layers~\cite{baenitz2021} induce nonzero effective magnetic fields in the Ag planes where also the excess Cr impurities are located. Due to this, time reversal symmetry is broken at the impurity sites and the Kondo coupling with the conduction electrons is strongly suppressed. At $T_\text N$, static magnetic order sets in: in the $ab$ plane, the Cr spins order in a cycloidal fashion, however along the $c$ direction, Cr spins are ordered antiferromagnetically. This static antiferromagnetic arrangement along $c$ effectively cancels the induced magnetic fields in the Ag plus Cr impurities planes intercalated between the \ce{CrSe2} layers, therefore allowing for the Kondo effect to emerge on the same energy scale as the antiferromagnetism.

The insights we gained here enable the study of systems from across the Doniach phase diagram~\cite{hewson1993} via impurity physics, even within the Kondo limit. It would be interesting to further explore the intricate interplay of magnetic interactions, such as the Ruderman-Kittel-Kasuya-Yosida interaction and various forms of magnetic scattering, across the phase diagram. Furthermore, it is important to contextualize our data within the Kondo framework, the concurrence of $T_\text N$ and $T_\text K$  is only possible in the Kondo single-ion model. In the traditional scenarios of competition between magnetic order and Kondo screening, such as the Kondo lattice model~\cite{hewson1993} and the Kondo necklace model~\cite{Thalmeier2006}, the appearance of magnetic order suppresses the Kondo effect.

To summarize, we have measured in detail the electronic transport properties of \ce{AgCrSe2} at varying magnetic fields and temperatures. We observed a highly anisotropic resistivity, characteristic of a layered material. Careful analysis of the in-plane resistivity revealed an impurity Kondo behaviour with a characteristic temperature of $T_\text K\approx34\,\rm K$ in zero external magnetic field and a Kondo breakdown field consistent with both the estimated Kondo temperature and an impurity spin $S=3/2$.

Magnetoresistance measurements confirmed the Kondo like transport with a characteristic temperature of $T_\text K\approx32\,\rm K$, and the excellent agreement with the theoretically expected behaviour allows us to conclude that the magnetic impurities (i) are the origin of the Kondo scattering and (ii) have spin $S=3/2$ in accordance with the value obtained from the Kondo breakdown field.

Finally we have discussed the relation between the antiferromagnetic order at $T_\text N$ and the Kondo impurity scattering. To the best of our knowledge, our results uncovered a unique feature of \ce{AgCrSe2}: the Néel temperature and the Kondo temperature coincide. We present a consistent physical picture of the mechanism behind that observation based on the location of the impurities and the cancellation of effective internal magnetic fields at these locations due to the onset of magnetic order, needed for the emergence of the impurity Kondo physics we observe in our magnetotransport measurements. Furthermore, our work would stimulate future studies to tune the interplay between Kondo screening and antiferromagnetic correlations on similar layered compounds. 

\section*{Methods}

\subsection*{Device preparation}

Single crystals of \ce{AgCrSe2} were grown via chemical vapor transport techniques, following the methodologies outlined in Baenitz et al~\cite{baenitz2021}. A device was fabricated with a single crystal of length $\approx$ 2 mm, width $\approx$ 1 mm and a thickness $\approx$ 50 $\mu$m, the contacts were deposited by sputtering gold in the crystal whilst it was covered by a shadow mask.

\subsection*{Transport measurements}

Transport measurements were carried out in Helium-4 cryostats, equipped with a 9 T and a 14 T magnet, respectively. SR830 lock-in amplifiers were used to perform four-point resistance measurements. An alternating current was applied between the source and drain contacts, and the transverse and longitudinal voltages were measured simultaneously. We used the standard procedure of symmetrizing the longitudinal resistivity, ${\rho_{\text{ab}}(H_\rightarrow) + \rho_{\text{ab}}(-H_\leftarrow)}/{2} = \rho_\text {ab}$, where $H_{\rightarrow,\leftarrow}$ refers to the field sweep direction.

\subsection*{Magnetization measurements}

Magnetization measurements were performed using a commercial vibrating sample magnetometer (VSM, Quantum Design).

\subsection*{X-ray spectroscopy}

The stoichiometry of \ce{AgCrSe2} was determined by EDX (energy dispersive x-ray spectroscopy)~\cite{heinrich1968} analysis on a metallographically mounted and polished crystal. The analyses were performed on a scanning electron microscope (Jeol JSM 7800 F) with attached EDX (energy dispersive X- ray spectroscopy) system (Bruker Quantax 400). The stoichiometry results were confirmed and measured with higher precision making use of wavelength dispersive x-ray measurements.\\
\\

\section{Data Availability}

All relevant data presented in the Manuscript are available from the corresponding author upon reasonable request.

\section{Code Availability}

All code relevant in analysing the data presented in the Manuscript are available from the corresponding author upon reasonable request.

\section{References}

%\bibliography{a.bib}

\begin{thebibliography}{40}%
\makeatletter
\providecommand \@ifxundefined [1]{%
 \@ifx{#1\undefined}
}%
\providecommand \@ifnum [1]{%
 \ifnum #1\expandafter \@firstoftwo
 \else \expandafter \@secondoftwo
 \fi
}%
\providecommand \@ifx [1]{%
 \ifx #1\expandafter \@firstoftwo
 \else \expandafter \@secondoftwo
 \fi
}%
\providecommand \natexlab [1]{#1}%
\providecommand \enquote  [1]{``#1''}%
\providecommand \bibnamefont  [1]{#1}%
\providecommand \bibfnamefont [1]{#1}%
\providecommand \citenamefont [1]{#1}%
\providecommand \href@noop [0]{\@secondoftwo}%
\providecommand \href [0]{\begingroup \@sanitize@url \@href}%
\providecommand \@href[1]{\@@startlink{#1}\@@href}%
\providecommand \@@href[1]{\endgroup#1\@@endlink}%
\providecommand \@sanitize@url [0]{\catcode `\\12\catcode `\$12\catcode
  `\&12\catcode `\#12\catcode `\^12\catcode `\_12\catcode `\%12\relax}%
\providecommand \@@startlink[1]{}%
\providecommand \@@endlink[0]{}%
\providecommand \url  [0]{\begingroup\@sanitize@url \@url }%
\providecommand \@url [1]{\endgroup\@href {#1}{\urlprefix }}%
\providecommand \urlprefix  [0]{URL }%
\providecommand \Eprint [0]{\href }%
\providecommand \doibase [0]{https://doi.org/}%
\providecommand \selectlanguage [0]{\@gobble}%
\providecommand \bibinfo  [0]{\@secondoftwo}%
\providecommand \bibfield  [0]{\@secondoftwo}%
\providecommand \translation [1]{[#1]}%
\providecommand \BibitemOpen [0]{}%
\providecommand \bibitemStop [0]{}%
\providecommand \bibitemNoStop [0]{.\EOS\space}%
\providecommand \EOS [0]{\spacefactor3000\relax}%
\providecommand \BibitemShut  [1]{\csname bibitem#1\endcsname}%
\let\auto@bib@innerbib\@empty
%</preamble>
\bibitem [{\citenamefont {Damay}\ \emph {et~al.}(2013)\citenamefont {Damay},
  \citenamefont {Petit}, \citenamefont {Braendlein}, \citenamefont {Rols},
  \citenamefont {Ollivier}, \citenamefont {Martin},\ and\ \citenamefont
  {Maignan}}]{damay:13}%
  \BibitemOpen
  \bibfield  {author} {\bibinfo {author} {\bibfnamefont {F.}~\bibnamefont
  {Damay}}, \bibinfo {author} {\bibfnamefont {S.}~\bibnamefont {Petit}},
  \bibinfo {author} {\bibfnamefont {M.}~\bibnamefont {Braendlein}}, \bibinfo
  {author} {\bibfnamefont {S.}~\bibnamefont {Rols}}, \bibinfo {author}
  {\bibfnamefont {J.}~\bibnamefont {Ollivier}}, \bibinfo {author}
  {\bibfnamefont {C.}~\bibnamefont {Martin}},\ and\ \bibinfo {author}
  {\bibfnamefont {A.}~\bibnamefont {Maignan}},\ }\bibfield  {title} {\bibinfo
  {title} {Spin dynamics in the unconventional multiferroic {AgCrS$_2$}},\
  }\bibfield  {journal} {\bibinfo  {journal} {Physical Review B}\ }\textbf
  {\bibinfo {volume} {87}},\ \href {https://doi.org/10.1103/PhysRevB.87.134413}
  {10.1103/PhysRevB.87.134413} (\bibinfo {year} {2013})\BibitemShut {NoStop}%
\bibitem [{\citenamefont {Matsuda}\ \emph {et~al.}(2020)\citenamefont
  {Matsuda}, \citenamefont {Dissanayake}, \citenamefont {Yoshida},
  \citenamefont {Isobe},\ and\ \citenamefont {Stone}}]{matsuda:20}%
  \BibitemOpen
  \bibfield  {author} {\bibinfo {author} {\bibfnamefont {M.}~\bibnamefont
  {Matsuda}}, \bibinfo {author} {\bibfnamefont {S.~E.}\ \bibnamefont
  {Dissanayake}}, \bibinfo {author} {\bibfnamefont {H.~K.}\ \bibnamefont
  {Yoshida}}, \bibinfo {author} {\bibfnamefont {M.}~\bibnamefont {Isobe}},\
  and\ \bibinfo {author} {\bibfnamefont {M.~B.}\ \bibnamefont {Stone}},\
  }\bibfield  {title} {\bibinfo {title} {Magnetic excitations affected by
  spin-lattice coupling in the {$S=3/2$} triangular lattice antiferromagnet
  {Ag$_2$CrO$_2$}},\ }\href {https://doi.org/10.1103/PhysRevB.102.214411}
  {\bibfield  {journal} {\bibinfo  {journal} {Phys. Rev. B}\ }\textbf {\bibinfo
  {volume} {102}},\ \bibinfo {pages} {214411} (\bibinfo {year}
  {2020})}\BibitemShut {NoStop}%
\bibitem [{\citenamefont {Baibich}\ \emph {et~al.}(1988)\citenamefont
  {Baibich}, \citenamefont {Broto}, \citenamefont {Fert}, \citenamefont
  {Van~Dau}, \citenamefont {Petroff}, \citenamefont {Etienne}, \citenamefont
  {Creuzet}, \citenamefont {Friederich},\ and\ \citenamefont
  {Chazelas}}]{PhysRevLett.61.2472}%
  \BibitemOpen
  \bibfield  {author} {\bibinfo {author} {\bibfnamefont {M.~N.}\ \bibnamefont
  {Baibich}}, \bibinfo {author} {\bibfnamefont {J.~M.}\ \bibnamefont {Broto}},
  \bibinfo {author} {\bibfnamefont {A.}~\bibnamefont {Fert}}, \bibinfo {author}
  {\bibfnamefont {F.~N.}\ \bibnamefont {Van~Dau}}, \bibinfo {author}
  {\bibfnamefont {F.}~\bibnamefont {Petroff}}, \bibinfo {author} {\bibfnamefont
  {P.}~\bibnamefont {Etienne}}, \bibinfo {author} {\bibfnamefont
  {G.}~\bibnamefont {Creuzet}}, \bibinfo {author} {\bibfnamefont
  {A.}~\bibnamefont {Friederich}},\ and\ \bibinfo {author} {\bibfnamefont
  {J.}~\bibnamefont {Chazelas}},\ }\bibfield  {title} {\bibinfo {title} {Giant
  magnetoresistance of (001)\uppercase{F}e/(001)\uppercase{C}r magnetic
  superlattices},\ }\href {https://doi.org/10.1103/PhysRevLett.61.2472}
  {\bibfield  {journal} {\bibinfo  {journal} {Phys. Rev. Lett.}\ }\textbf
  {\bibinfo {volume} {61}},\ \bibinfo {pages} {2472} (\bibinfo {year}
  {1988})}\BibitemShut {NoStop}%
\bibitem [{\citenamefont {Ideue}\ \emph {et~al.}(2020)\citenamefont {Ideue},
  \citenamefont {Koshikawa}, \citenamefont {Namiki}, \citenamefont {Sasagawa},\
  and\ \citenamefont {Iwasa}}]{Ideue2020}%
  \BibitemOpen
  \bibfield  {author} {\bibinfo {author} {\bibfnamefont {T.}~\bibnamefont
  {Ideue}}, \bibinfo {author} {\bibfnamefont {S.}~\bibnamefont {Koshikawa}},
  \bibinfo {author} {\bibfnamefont {H.}~\bibnamefont {Namiki}}, \bibinfo
  {author} {\bibfnamefont {T.}~\bibnamefont {Sasagawa}},\ and\ \bibinfo
  {author} {\bibfnamefont {Y.}~\bibnamefont {Iwasa}},\ }\bibfield  {title}
  {\bibinfo {title} {Giant nonreciprocal magnetotransport in bulk trigonal
  superconductor {PbTaSe}$_2$},\ }\href
  {https://doi.org/10.1103/PhysRevResearch.2.042046} {\bibfield  {journal}
  {\bibinfo  {journal} {Phys. Rev. Res.}\ }\textbf {\bibinfo {volume} {2}},\
  \bibinfo {pages} {042046} (\bibinfo {year} {2020})}\BibitemShut {NoStop}%
\bibitem [{\citenamefont {Fert}\ \emph {et~al.}(2017)\citenamefont {Fert},
  \citenamefont {Reyren},\ and\ \citenamefont {Cros}}]{fert2017magnetic}%
  \BibitemOpen
  \bibfield  {author} {\bibinfo {author} {\bibfnamefont {A.}~\bibnamefont
  {Fert}}, \bibinfo {author} {\bibfnamefont {N.}~\bibnamefont {Reyren}},\ and\
  \bibinfo {author} {\bibfnamefont {V.}~\bibnamefont {Cros}},\ }\bibfield
  {title} {\bibinfo {title} {Magnetic skyrmions: advances in physics and
  potential applications},\ }\href
  {https://doi.org/https://doi.org/10.1038/natrevmats.2017.31} {\bibfield
  {journal} {\bibinfo  {journal} {Nature Reviews Materials}\ }\textbf {\bibinfo
  {volume} {2}},\ \bibinfo {pages} {1} (\bibinfo {year} {2017})}\BibitemShut
  {NoStop}%
\bibitem [{\citenamefont {Wu}\ \emph {et~al.}(2016)\citenamefont {Wu},
  \citenamefont {Huang}, \citenamefont {Feng}, \citenamefont {Li},
  \citenamefont {Chen}, \citenamefont {Zhang},\ and\ \citenamefont
  {He}}]{Wu2016}%
  \BibitemOpen
  \bibfield  {author} {\bibinfo {author} {\bibfnamefont {D.}~\bibnamefont
  {Wu}}, \bibinfo {author} {\bibfnamefont {S.}~\bibnamefont {Huang}}, \bibinfo
  {author} {\bibfnamefont {D.}~\bibnamefont {Feng}}, \bibinfo {author}
  {\bibfnamefont {B.}~\bibnamefont {Li}}, \bibinfo {author} {\bibfnamefont
  {Y.}~\bibnamefont {Chen}}, \bibinfo {author} {\bibfnamefont {J.}~\bibnamefont
  {Zhang}},\ and\ \bibinfo {author} {\bibfnamefont {J.}~\bibnamefont {He}},\
  }\bibfield  {title} {\bibinfo {title} {Revisiting {AgCrSe}$_2$ as a promising
  thermoelectric material},\ }\href {https://doi.org/10.1039/C6CP04791B}
  {\bibfield  {journal} {\bibinfo  {journal} {Phys. Chem. Chem. Phys.}\
  }\textbf {\bibinfo {volume} {18}},\ \bibinfo {pages} {23872} (\bibinfo {year}
  {2016})}\BibitemShut {NoStop}%
\bibitem [{\citenamefont {Shiomi}\ \emph {et~al.}(2018)\citenamefont {Shiomi},
  \citenamefont {Akiba}, \citenamefont {Takahashi},\ and\ \citenamefont
  {Ishiwata}}]{Shiomi2018}%
  \BibitemOpen
  \bibfield  {author} {\bibinfo {author} {\bibfnamefont {Y.}~\bibnamefont
  {Shiomi}}, \bibinfo {author} {\bibfnamefont {T.}~\bibnamefont {Akiba}},
  \bibinfo {author} {\bibfnamefont {H.}~\bibnamefont {Takahashi}},\ and\
  \bibinfo {author} {\bibfnamefont {S.}~\bibnamefont {Ishiwata}},\ }\bibfield
  {title} {\bibinfo {title} {Giant piezoelectric response in superionic polar
  semiconductor},\ }\href
  {https://doi.org/https://doi.org/10.1002/aelm.201800174} {\bibfield
  {journal} {\bibinfo  {journal} {Advanced Electronic Materials}\ }\textbf
  {\bibinfo {volume} {4}},\ \bibinfo {pages} {1800174} (\bibinfo {year}
  {2018})}\BibitemShut {NoStop}%
\bibitem [{\citenamefont {Hua}\ \emph {et~al.}(2021)\citenamefont {Hua},
  \citenamefont {Bai}, \citenamefont {Wang}, \citenamefont {Wu}, \citenamefont
  {Cui}, \citenamefont {Sun},\ and\ \citenamefont {Xiao}}]{hua2021}%
  \BibitemOpen
  \bibfield  {author} {\bibinfo {author} {\bibfnamefont {Y.}~\bibnamefont
  {Hua}}, \bibinfo {author} {\bibfnamefont {W.}~\bibnamefont {Bai}}, \bibinfo
  {author} {\bibfnamefont {S.}~\bibnamefont {Wang}}, \bibinfo {author}
  {\bibfnamefont {Y.}~\bibnamefont {Wu}}, \bibinfo {author} {\bibfnamefont
  {S.}~\bibnamefont {Cui}}, \bibinfo {author} {\bibfnamefont {Z.}~\bibnamefont
  {Sun}},\ and\ \bibinfo {author} {\bibfnamefont {C.}~\bibnamefont {Xiao}},\
  }\bibfield  {title} {\bibinfo {title} {Tuning the electric transport behavior
  of {AgCrSe}$_2$ by intrinsic defects},\ }\href
  {https://doi.org/https://doi.org/10.1007/s11426-021-1071-4} {\bibfield
  {journal} {\bibinfo  {journal} {Science China Chemistry}\ }\textbf {\bibinfo
  {volume} {64}},\ \bibinfo {pages} {1970} (\bibinfo {year}
  {2021})}\BibitemShut {NoStop}%
\bibitem [{\citenamefont {Takahashi}\ \emph {et~al.}(2022)\citenamefont
  {Takahashi}, \citenamefont {Akiba}, \citenamefont {Mayo}, \citenamefont
  {Akiba}, \citenamefont {Miyake}, \citenamefont {Tokunaga}, \citenamefont
  {Mori}, \citenamefont {Arita},\ and\ \citenamefont
  {Ishiwata}}]{takahashi2022}%
  \BibitemOpen
  \bibfield  {author} {\bibinfo {author} {\bibfnamefont {H.}~\bibnamefont
  {Takahashi}}, \bibinfo {author} {\bibfnamefont {T.}~\bibnamefont {Akiba}},
  \bibinfo {author} {\bibfnamefont {A.~H.}\ \bibnamefont {Mayo}}, \bibinfo
  {author} {\bibfnamefont {K.}~\bibnamefont {Akiba}}, \bibinfo {author}
  {\bibfnamefont {A.}~\bibnamefont {Miyake}}, \bibinfo {author} {\bibfnamefont
  {M.}~\bibnamefont {Tokunaga}}, \bibinfo {author} {\bibfnamefont
  {H.}~\bibnamefont {Mori}}, \bibinfo {author} {\bibfnamefont {R.}~\bibnamefont
  {Arita}},\ and\ \bibinfo {author} {\bibfnamefont {S.}~\bibnamefont
  {Ishiwata}},\ }\bibfield  {title} {\bibinfo {title} {Spin-orbit-derived giant
  magnetoresistance in a layered magnetic semiconductor {AgCrSe}$_2$},\ }\href
  {https://doi.org/10.1103/PhysRevMaterials.6.054602} {\bibfield  {journal}
  {\bibinfo  {journal} {Phys. Rev. Mater.}\ }\textbf {\bibinfo {volume} {6}},\
  \bibinfo {pages} {054602} (\bibinfo {year} {2022})}\BibitemShut {NoStop}%
\bibitem [{\citenamefont {Kim}\ \emph {et~al.}(2023)\citenamefont {Kim},
  \citenamefont {Zhu}, \citenamefont {Piva}, \citenamefont {Schmidt},
  \citenamefont {Fartab}, \citenamefont {Mackenzie}, \citenamefont {Baenitz},
  \citenamefont {Nicklas}, \citenamefont {Rosner}, \citenamefont {Cook},
  \citenamefont {González-Hernández}, \citenamefont {Šmejkal},\ and\
  \citenamefont {Zhang}}]{kim2023}%
  \BibitemOpen
  \bibfield  {author} {\bibinfo {author} {\bibfnamefont {S.-J.}\ \bibnamefont
  {Kim}}, \bibinfo {author} {\bibfnamefont {J.}~\bibnamefont {Zhu}}, \bibinfo
  {author} {\bibfnamefont {M.~M.}\ \bibnamefont {Piva}}, \bibinfo {author}
  {\bibfnamefont {M.}~\bibnamefont {Schmidt}}, \bibinfo {author} {\bibfnamefont
  {D.}~\bibnamefont {Fartab}}, \bibinfo {author} {\bibfnamefont {A.~P.}\
  \bibnamefont {Mackenzie}}, \bibinfo {author} {\bibfnamefont {M.}~\bibnamefont
  {Baenitz}}, \bibinfo {author} {\bibfnamefont {M.}~\bibnamefont {Nicklas}},
  \bibinfo {author} {\bibfnamefont {H.}~\bibnamefont {Rosner}}, \bibinfo
  {author} {\bibfnamefont {A.~M.}\ \bibnamefont {Cook}}, \bibinfo {author}
  {\bibfnamefont {R.}~\bibnamefont {González-Hernández}}, \bibinfo {author}
  {\bibfnamefont {L.}~\bibnamefont {Šmejkal}},\ and\ \bibinfo {author}
  {\bibfnamefont {H.}~\bibnamefont {Zhang}},\ }\bibfield  {title} {\bibinfo
  {title} {Observation of the anomalous \uppercase{H}all effect in a layered
  polar semiconductor},\ }\href
  {https://doi.org/https://doi.org/10.1002/advs.202307306} {\bibfield
  {journal} {\bibinfo  {journal} {Advanced Science}\ ,\ \bibinfo {pages}
  {2307306}} (\bibinfo {year} {2023})}\BibitemShut {NoStop}%
\bibitem [{\citenamefont {Baenitz}\ \emph {et~al.}(2021)\citenamefont
  {Baenitz}, \citenamefont {Piva}, \citenamefont {Luther}, \citenamefont
  {Sichelschmidt}, \citenamefont {Ranjith}, \citenamefont
  {Dawczak-D\ifmmode~\mbox{\c{e}}\else \c{e}\fi{}bicki}, \citenamefont
  {Ajeesh}, \citenamefont {Kim}, \citenamefont {Siemann}, \citenamefont {Bigi},
  \citenamefont {Manuel}, \citenamefont {Khalyavin}, \citenamefont {Sokolov},
  \citenamefont {Mokhtari}, \citenamefont {Zhang}, \citenamefont {Yasuoka},
  \citenamefont {King}, \citenamefont {Vinai}, \citenamefont {Polewczyk},
  \citenamefont {Torelli}, \citenamefont {Wosnitza}, \citenamefont {Burkhardt},
  \citenamefont {Schmidt}, \citenamefont {Rosner}, \citenamefont {Wirth},
  \citenamefont {K\"uhne}, \citenamefont {Nicklas},\ and\ \citenamefont
  {Schmidt}}]{baenitz2021}%
  \BibitemOpen
  \bibfield  {author} {\bibinfo {author} {\bibfnamefont {M.}~\bibnamefont
  {Baenitz}}, \bibinfo {author} {\bibfnamefont {M.~M.}\ \bibnamefont {Piva}},
  \bibinfo {author} {\bibfnamefont {S.}~\bibnamefont {Luther}}, \bibinfo
  {author} {\bibfnamefont {J.}~\bibnamefont {Sichelschmidt}}, \bibinfo {author}
  {\bibfnamefont {K.~M.}\ \bibnamefont {Ranjith}}, \bibinfo {author}
  {\bibfnamefont {H.}~\bibnamefont {Dawczak-D\ifmmode~\mbox{\c{e}}\else
  \c{e}\fi{}bicki}}, \bibinfo {author} {\bibfnamefont {M.~O.}\ \bibnamefont
  {Ajeesh}}, \bibinfo {author} {\bibfnamefont {S.-J.}\ \bibnamefont {Kim}},
  \bibinfo {author} {\bibfnamefont {G.}~\bibnamefont {Siemann}}, \bibinfo
  {author} {\bibfnamefont {C.}~\bibnamefont {Bigi}}, \bibinfo {author}
  {\bibfnamefont {P.}~\bibnamefont {Manuel}}, \bibinfo {author} {\bibfnamefont
  {D.}~\bibnamefont {Khalyavin}}, \bibinfo {author} {\bibfnamefont {D.~A.}\
  \bibnamefont {Sokolov}}, \bibinfo {author} {\bibfnamefont {P.}~\bibnamefont
  {Mokhtari}}, \bibinfo {author} {\bibfnamefont {H.}~\bibnamefont {Zhang}},
  \bibinfo {author} {\bibfnamefont {H.}~\bibnamefont {Yasuoka}}, \bibinfo
  {author} {\bibfnamefont {P.~D.~C.}\ \bibnamefont {King}}, \bibinfo {author}
  {\bibfnamefont {G.}~\bibnamefont {Vinai}}, \bibinfo {author} {\bibfnamefont
  {V.}~\bibnamefont {Polewczyk}}, \bibinfo {author} {\bibfnamefont
  {P.}~\bibnamefont {Torelli}}, \bibinfo {author} {\bibfnamefont
  {J.}~\bibnamefont {Wosnitza}}, \bibinfo {author} {\bibfnamefont
  {U.}~\bibnamefont {Burkhardt}}, \bibinfo {author} {\bibfnamefont
  {B.}~\bibnamefont {Schmidt}}, \bibinfo {author} {\bibfnamefont
  {H.}~\bibnamefont {Rosner}}, \bibinfo {author} {\bibfnamefont
  {S.}~\bibnamefont {Wirth}}, \bibinfo {author} {\bibfnamefont
  {H.}~\bibnamefont {K\"uhne}}, \bibinfo {author} {\bibfnamefont
  {M.}~\bibnamefont {Nicklas}},\ and\ \bibinfo {author} {\bibfnamefont
  {M.}~\bibnamefont {Schmidt}},\ }\bibfield  {title} {\bibinfo {title} {Planar
  triangular {$S=3/2$} magnet {AgCrSe$_2$}: Magnetic frustration, short-range
  correlations, and field-tuned anisotropic cycloidal magnetic order},\ }\href
  {https://doi.org/10.1103/PhysRevB.104.134410} {\bibfield  {journal} {\bibinfo
   {journal} {Phys. Rev. B}\ }\textbf {\bibinfo {volume} {104}},\ \bibinfo
  {pages} {134410} (\bibinfo {year} {2021})}\BibitemShut {NoStop}%
\bibitem [{\citenamefont {Hewson}(1993)}]{hewson1993}%
  \BibitemOpen
  \bibfield  {author} {\bibinfo {author} {\bibfnamefont {A.~C.}\ \bibnamefont
  {Hewson}},\ }\href {https://doi.org/https://doi.org/10.1017/CBO9780511470752}
  {\emph {\bibinfo {title} {The Kondo Problem to Heavy Fermions}}}\ (\bibinfo
  {publisher} {Cambridge University Press},\ \bibinfo {address} {Cambridge},\
  \bibinfo {year} {1993})\BibitemShut {NoStop}%
\bibitem [{\citenamefont {Nagaosa}\ and\ \citenamefont
  {Lee}(1997)}]{Nagaosa1997}%
  \BibitemOpen
  \bibfield  {author} {\bibinfo {author} {\bibfnamefont {N.}~\bibnamefont
  {Nagaosa}}\ and\ \bibinfo {author} {\bibfnamefont {P.~A.}\ \bibnamefont
  {Lee}},\ }\bibfield  {title} {\bibinfo {title} {Kondo effect in high-
  ${T}_{c}$ cuprates},\ }\href {https://doi.org/10.1103/PhysRevLett.79.3755}
  {\bibfield  {journal} {\bibinfo  {journal} {Phys. Rev. Lett.}\ }\textbf
  {\bibinfo {volume} {79}},\ \bibinfo {pages} {3755} (\bibinfo {year}
  {1997})}\BibitemShut {NoStop}%
\bibitem [{\citenamefont {Scalapino}(2012)}]{RevModPhys.84.1383}%
  \BibitemOpen
  \bibfield  {author} {\bibinfo {author} {\bibfnamefont {D.~J.}\ \bibnamefont
  {Scalapino}},\ }\bibfield  {title} {\bibinfo {title} {A common thread: The
  pairing interaction for unconventional superconductors},\ }\href
  {https://doi.org/10.1103/RevModPhys.84.1383} {\bibfield  {journal} {\bibinfo
  {journal} {Rev. Mod. Phys.}\ }\textbf {\bibinfo {volume} {84}},\ \bibinfo
  {pages} {1383} (\bibinfo {year} {2012})}\BibitemShut {NoStop}%
\bibitem [{\citenamefont {Paschen}\ and\ \citenamefont
  {Si}(2021)}]{paschen2021quantum}%
  \BibitemOpen
  \bibfield  {author} {\bibinfo {author} {\bibfnamefont {S.}~\bibnamefont
  {Paschen}}\ and\ \bibinfo {author} {\bibfnamefont {Q.}~\bibnamefont {Si}},\
  }\bibfield  {title} {\bibinfo {title} {Quantum phases driven by strong
  correlations},\ }\href
  {https://doi.org/https://doi.org/10.1038/s42254-020-00262-6} {\bibfield
  {journal} {\bibinfo  {journal} {Nature Reviews Physics}\ }\textbf {\bibinfo
  {volume} {3}},\ \bibinfo {pages} {9} (\bibinfo {year} {2021})}\BibitemShut
  {NoStop}%
\bibitem [{\citenamefont {Li}\ \emph {et~al.}(2018)\citenamefont {Li},
  \citenamefont {Wang}, \citenamefont {Kawakita}, \citenamefont {Zhang},
  \citenamefont {Feygenson}, \citenamefont {Yu}, \citenamefont {Wu},
  \citenamefont {Ohara}, \citenamefont {Kikuchi}, \citenamefont {Shibata},
  \citenamefont {Yamada}, \citenamefont {Ning}, \citenamefont {Chen},
  \citenamefont {He}, \citenamefont {Vaknin}, \citenamefont {Wu}, \citenamefont
  {Nakajima},\ and\ \citenamefont {Kanatzidis}}]{li2018}%
  \BibitemOpen
  \bibfield  {author} {\bibinfo {author} {\bibfnamefont {B.}~\bibnamefont
  {Li}}, \bibinfo {author} {\bibfnamefont {H.}~\bibnamefont {Wang}}, \bibinfo
  {author} {\bibfnamefont {Y.}~\bibnamefont {Kawakita}}, \bibinfo {author}
  {\bibfnamefont {Q.}~\bibnamefont {Zhang}}, \bibinfo {author} {\bibfnamefont
  {M.}~\bibnamefont {Feygenson}}, \bibinfo {author} {\bibfnamefont {H.~L.}\
  \bibnamefont {Yu}}, \bibinfo {author} {\bibfnamefont {D.}~\bibnamefont {Wu}},
  \bibinfo {author} {\bibfnamefont {K.}~\bibnamefont {Ohara}}, \bibinfo
  {author} {\bibfnamefont {T.}~\bibnamefont {Kikuchi}}, \bibinfo {author}
  {\bibfnamefont {K.}~\bibnamefont {Shibata}}, \bibinfo {author} {\bibfnamefont
  {T.}~\bibnamefont {Yamada}}, \bibinfo {author} {\bibfnamefont {X.~K.}\
  \bibnamefont {Ning}}, \bibinfo {author} {\bibfnamefont {Y.}~\bibnamefont
  {Chen}}, \bibinfo {author} {\bibfnamefont {J.~Q.}\ \bibnamefont {He}},
  \bibinfo {author} {\bibfnamefont {D.}~\bibnamefont {Vaknin}}, \bibinfo
  {author} {\bibfnamefont {R.~Q.}\ \bibnamefont {Wu}}, \bibinfo {author}
  {\bibfnamefont {K.}~\bibnamefont {Nakajima}},\ and\ \bibinfo {author}
  {\bibfnamefont {M.~G.}\ \bibnamefont {Kanatzidis}},\ }\bibfield  {title}
  {\bibinfo {title} {Liquid-like thermal conduction in intercalated layered
  crystalline solids},\ }\href
  {https://doi.org/https://doi.org/10.1038/s41563-017-0004-2} {\bibfield
  {journal} {\bibinfo  {journal} {Nature Materials}\ }\textbf {\bibinfo
  {volume} {17}},\ \bibinfo {pages} {226} (\bibinfo {year} {2018})}\BibitemShut
  {NoStop}%
\bibitem [{\citenamefont {Wang}\ and\ \citenamefont {Chen}(2020)}]{wang2020}%
  \BibitemOpen
  \bibfield  {author} {\bibinfo {author} {\bibfnamefont {C.}~\bibnamefont
  {Wang}}\ and\ \bibinfo {author} {\bibfnamefont {Y.}~\bibnamefont {Chen}},\
  }\bibfield  {title} {\bibinfo {title} {Highly selective phonon diffusive
  scattering in superionic layered {AgCrSe}$_2$},\ }\href
  {https://doi.org/https://doi.org/10.1038/s41524-020-0295-8} {\bibfield
  {journal} {\bibinfo  {journal} {npj Computational Materials}\ }\textbf
  {\bibinfo {volume} {6}},\ \bibinfo {pages} {26} (\bibinfo {year}
  {2020})}\BibitemShut {NoStop}%
\bibitem [{\citenamefont {Siemann}\ \emph {et~al.}(2023)\citenamefont
  {Siemann}, \citenamefont {Kim}, \citenamefont {Morales}, \citenamefont
  {Murgatroyd}, \citenamefont {Zivanovic}, \citenamefont {Edwards},
  \citenamefont {Markovi{\'c}}, \citenamefont {Mazzola}, \citenamefont
  {Trzaska}, \citenamefont {Clark}, \citenamefont {Bigi}, \citenamefont
  {Zhang}, \citenamefont {Achinuq}, \citenamefont {Hesjedal}, \citenamefont
  {Watson}, \citenamefont {Kim}, \citenamefont {Bencok}, \citenamefont {van~der
  Laan}, \citenamefont {Polley}, \citenamefont {Leandersson}, \citenamefont
  {Fedderwitz}, \citenamefont {Ali}, \citenamefont {Balasubramanian},
  \citenamefont {Schmidt}, \citenamefont {Baenitz}, \citenamefont {Rosner},\
  and\ \citenamefont {King}}]{gesa2023}%
  \BibitemOpen
  \bibfield  {author} {\bibinfo {author} {\bibfnamefont {G.-R.}\ \bibnamefont
  {Siemann}}, \bibinfo {author} {\bibfnamefont {S.-J.}\ \bibnamefont {Kim}},
  \bibinfo {author} {\bibfnamefont {E.~A.}\ \bibnamefont {Morales}}, \bibinfo
  {author} {\bibfnamefont {P.~A.~E.}\ \bibnamefont {Murgatroyd}}, \bibinfo
  {author} {\bibfnamefont {A.}~\bibnamefont {Zivanovic}}, \bibinfo {author}
  {\bibfnamefont {B.}~\bibnamefont {Edwards}}, \bibinfo {author} {\bibfnamefont
  {I.}~\bibnamefont {Markovi{\'c}}}, \bibinfo {author} {\bibfnamefont
  {F.}~\bibnamefont {Mazzola}}, \bibinfo {author} {\bibfnamefont
  {L.}~\bibnamefont {Trzaska}}, \bibinfo {author} {\bibfnamefont {O.~J.}\
  \bibnamefont {Clark}}, \bibinfo {author} {\bibfnamefont {C.}~\bibnamefont
  {Bigi}}, \bibinfo {author} {\bibfnamefont {H.}~\bibnamefont {Zhang}},
  \bibinfo {author} {\bibfnamefont {B.}~\bibnamefont {Achinuq}}, \bibinfo
  {author} {\bibfnamefont {T.}~\bibnamefont {Hesjedal}}, \bibinfo {author}
  {\bibfnamefont {M.~D.}\ \bibnamefont {Watson}}, \bibinfo {author}
  {\bibfnamefont {T.~K.}\ \bibnamefont {Kim}}, \bibinfo {author} {\bibfnamefont
  {P.}~\bibnamefont {Bencok}}, \bibinfo {author} {\bibfnamefont
  {G.}~\bibnamefont {van~der Laan}}, \bibinfo {author} {\bibfnamefont {C.~M.}\
  \bibnamefont {Polley}}, \bibinfo {author} {\bibfnamefont {M.}~\bibnamefont
  {Leandersson}}, \bibinfo {author} {\bibfnamefont {H.}~\bibnamefont
  {Fedderwitz}}, \bibinfo {author} {\bibfnamefont {K.}~\bibnamefont {Ali}},
  \bibinfo {author} {\bibfnamefont {T.}~\bibnamefont {Balasubramanian}},
  \bibinfo {author} {\bibfnamefont {M.}~\bibnamefont {Schmidt}}, \bibinfo
  {author} {\bibfnamefont {M.}~\bibnamefont {Baenitz}}, \bibinfo {author}
  {\bibfnamefont {H.}~\bibnamefont {Rosner}},\ and\ \bibinfo {author}
  {\bibfnamefont {P.~D.~C.}\ \bibnamefont {King}},\ }\bibfield  {title}
  {\bibinfo {title} {Spin-orbit coupled spin-polarised hole gas at the
  {CrSe}$_2$-terminated surface of {AgCrSe}$_2$},\ }\href
  {https://doi.org/10.1038/s41535-023-00593-4} {\bibfield  {journal} {\bibinfo
  {journal} {npj Quantum Materials}\ }\textbf {\bibinfo {volume} {8}},\
  \bibinfo {pages} {61} (\bibinfo {year} {2023})}\BibitemShut {NoStop}%
\bibitem [{\citenamefont {Mackenzie}(2017)}]{Mackenzie2017}%
  \BibitemOpen
  \bibfield  {author} {\bibinfo {author} {\bibfnamefont {A.~P.}\ \bibnamefont
  {Mackenzie}},\ }\bibfield  {title} {\bibinfo {title} {The properties of
  ultrapure delafossite metals},\ }\href
  {https://doi.org/10.1088/1361-6633/aa50e5} {\bibfield  {journal} {\bibinfo
  {journal} {Reports on Progress in Physics}\ }\textbf {\bibinfo {volume}
  {80}},\ \bibinfo {pages} {032501} (\bibinfo {year} {2017})}\BibitemShut
  {NoStop}%
\bibitem [{\citenamefont {Fisher}(1962)}]{Fisher1962}%
  \BibitemOpen
  \bibfield  {author} {\bibinfo {author} {\bibfnamefont {M.~E.}\ \bibnamefont
  {Fisher}},\ }\bibfield  {title} {\bibinfo {title} {Relation between the
  specific heat and susceptibility of an antiferromagnet},\ }\href
  {https://doi.org/10.1080/14786436208213705} {\bibfield  {journal} {\bibinfo
  {journal} {The Philosophical Magazine: A Journal of Theoretical Experimental
  and Applied Physics}\ }\textbf {\bibinfo {volume} {7}},\ \bibinfo {pages}
  {1731} (\bibinfo {year} {1962})}\BibitemShut {NoStop}%
\bibitem [{\citenamefont {van~der Pauw}(1991)}]{vdp}%
  \BibitemOpen
  \bibfield  {author} {\bibinfo {author} {\bibfnamefont {L.~J.}\ \bibnamefont
  {van~der Pauw}},\ }\bibinfo {title} {A method of measuring specific
  resistivity and {H}all effects of discs of arbitrary shape},\ in\ \href
  {https://doi.org/10.1142/9789814503464_0017} {\emph {\bibinfo {booktitle}
  {Semiconductor Devices: Pioneering Papers}}}\ (\bibinfo {year} {1991})\ pp.\
  \bibinfo {pages} {174--182}\BibitemShut {NoStop}%
\bibitem [{\citenamefont {Zhang}\ \emph {et~al.}(2019)\citenamefont {Zhang},
  \citenamefont {Berthod}, \citenamefont {Berger}, \citenamefont {Giamarchi},\
  and\ \citenamefont {Morpurgo}}]{zhang2019band}%
  \BibitemOpen
  \bibfield  {author} {\bibinfo {author} {\bibfnamefont {H.}~\bibnamefont
  {Zhang}}, \bibinfo {author} {\bibfnamefont {C.}~\bibnamefont {Berthod}},
  \bibinfo {author} {\bibfnamefont {H.}~\bibnamefont {Berger}}, \bibinfo
  {author} {\bibfnamefont {T.}~\bibnamefont {Giamarchi}},\ and\ \bibinfo
  {author} {\bibfnamefont {A.~F.}\ \bibnamefont {Morpurgo}},\ }\bibfield
  {title} {\bibinfo {title} {Band filling and cross quantum capacitance in
  ion-gated semiconducting transition metal dichalcogenide monolayers},\ }\href
  {https://doi.org/http://dx.doi.org/10.1021/acs.nanolett.9b03667} {\bibfield
  {journal} {\bibinfo  {journal} {Nano letters}\ }\textbf {\bibinfo {volume}
  {19}},\ \bibinfo {pages} {8836} (\bibinfo {year} {2019})}\BibitemShut
  {NoStop}%
\bibitem [{\citenamefont {Valla}\ \emph {et~al.}(2002)\citenamefont {Valla},
  \citenamefont {Johnson}, \citenamefont {Yusof}, \citenamefont {Wells},
  \citenamefont {Li}, \citenamefont {Loureiro}, \citenamefont {Cava},
  \citenamefont {Mikami}, \citenamefont {Mori}, \citenamefont {Yoshimura},\
  and\ \citenamefont {Sasaki}}]{valla2002}%
  \BibitemOpen
  \bibfield  {author} {\bibinfo {author} {\bibfnamefont {T.}~\bibnamefont
  {Valla}}, \bibinfo {author} {\bibfnamefont {P.~D.}\ \bibnamefont {Johnson}},
  \bibinfo {author} {\bibfnamefont {Z.}~\bibnamefont {Yusof}}, \bibinfo
  {author} {\bibfnamefont {B.}~\bibnamefont {Wells}}, \bibinfo {author}
  {\bibfnamefont {Q.}~\bibnamefont {Li}}, \bibinfo {author} {\bibfnamefont
  {S.~M.}\ \bibnamefont {Loureiro}}, \bibinfo {author} {\bibfnamefont {R.~J.}\
  \bibnamefont {Cava}}, \bibinfo {author} {\bibfnamefont {M.}~\bibnamefont
  {Mikami}}, \bibinfo {author} {\bibfnamefont {Y.}~\bibnamefont {Mori}},
  \bibinfo {author} {\bibfnamefont {M.}~\bibnamefont {Yoshimura}},\ and\
  \bibinfo {author} {\bibfnamefont {T.}~\bibnamefont {Sasaki}},\ }\bibfield
  {title} {\bibinfo {title} {Coherence--incoherence and dimensional crossover
  in layered strongly correlated metals},\ }\href
  {https://doi.org/https://doi.org/10.1038/nature00774} {\bibfield  {journal}
  {\bibinfo  {journal} {Nature}\ }\textbf {\bibinfo {volume} {417}},\ \bibinfo
  {pages} {627} (\bibinfo {year} {2002})}\BibitemShut {NoStop}%
\bibitem [{\citenamefont {Yano}\ and\ \citenamefont
  {Sasagawa}(2016)}]{yano2016}%
  \BibitemOpen
  \bibfield  {author} {\bibinfo {author} {\bibfnamefont {R.}~\bibnamefont
  {Yano}}\ and\ \bibinfo {author} {\bibfnamefont {T.}~\bibnamefont
  {Sasagawa}},\ }\bibfield  {title} {\bibinfo {title} {Crystal growth and
  intrinsic properties of {ACrX}$_2$ ({A} = {Cu}, {Ag}; {X = S}, {Se}) without
  a secondary phase},\ }\href
  {https://doi.org/https://doi.org/10.1021/acs.cgd.6b00037} {\bibfield
  {journal} {\bibinfo  {journal} {Crystal Growth \& Design}\ }\textbf {\bibinfo
  {volume} {16}},\ \bibinfo {pages} {5618} (\bibinfo {year}
  {2016})}\BibitemShut {NoStop}%
\bibitem [{\citenamefont {Kondo}(1964)}]{kondo1964}%
  \BibitemOpen
  \bibfield  {author} {\bibinfo {author} {\bibfnamefont {J.}~\bibnamefont
  {Kondo}},\ }\bibfield  {title} {\bibinfo {title} {Resistance minimum in
  dilute magnetic alloys},\ }\href
  {https://api.semanticscholar.org/CorpusID:122790965} {\bibfield  {journal}
  {\bibinfo  {journal} {Progress of Theoretical Physics}\ }\textbf {\bibinfo
  {volume} {32}},\ \bibinfo {pages} {37} (\bibinfo {year} {1964})}\BibitemShut
  {NoStop}%
\bibitem [{\citenamefont {Abrikosov}(1965)}]{abrikosov1965}%
  \BibitemOpen
  \bibfield  {author} {\bibinfo {author} {\bibfnamefont {A.~A.}\ \bibnamefont
  {Abrikosov}},\ }\bibfield  {title} {\bibinfo {title} {Electron scattering on
  magnetic impurities in metals and anomalous resistivity effects},\ }\href
  {https://doi.org/10.1103/PhysicsPhysiqueFizika.2.5} {\bibfield  {journal}
  {\bibinfo  {journal} {Physics Physique Fizika}\ }\textbf {\bibinfo {volume}
  {2}},\ \bibinfo {pages} {5} (\bibinfo {year} {1965})}\BibitemShut {NoStop}%
\bibitem [{\citenamefont {Suhl}(1965)}]{suhl1965}%
  \BibitemOpen
  \bibfield  {author} {\bibinfo {author} {\bibfnamefont {H.}~\bibnamefont
  {Suhl}},\ }\bibfield  {title} {\bibinfo {title} {Dispersion theory of the
  kondo effect},\ }\href {https://doi.org/10.1103/PhysRev.138.A515} {\bibfield
  {journal} {\bibinfo  {journal} {Phys. Rev.}\ }\textbf {\bibinfo {volume}
  {138}},\ \bibinfo {pages} {A515} (\bibinfo {year} {1965})}\BibitemShut
  {NoStop}%
\bibitem [{\citenamefont {Zhu}\ \emph {et~al.}(2016)\citenamefont {Zhu},
  \citenamefont {Nie}, \citenamefont {Xiong}, \citenamefont {Schlottmann},\
  and\ \citenamefont {Zhao}}]{zhu2016}%
  \BibitemOpen
  \bibfield  {author} {\bibinfo {author} {\bibfnamefont {L.~J.}\ \bibnamefont
  {Zhu}}, \bibinfo {author} {\bibfnamefont {S.~H.}\ \bibnamefont {Nie}},
  \bibinfo {author} {\bibfnamefont {P.}~\bibnamefont {Xiong}}, \bibinfo
  {author} {\bibfnamefont {P.}~\bibnamefont {Schlottmann}},\ and\ \bibinfo
  {author} {\bibfnamefont {J.~H.}\ \bibnamefont {Zhao}},\ }\bibfield  {title}
  {\bibinfo {title} {Orbital two-channel kondo effect in epitaxial
  ferromagnetic {L}10-{MnAl} films},\ }\href
  {https://doi.org/https://doi.org/10.1038/ncomms10817} {\bibfield  {journal}
  {\bibinfo  {journal} {Nature Communications}\ }\textbf {\bibinfo {volume}
  {7}},\ \bibinfo {pages} {10817} (\bibinfo {year} {2016})}\BibitemShut
  {NoStop}%
\bibitem [{\citenamefont {Khadka}\ \emph {et~al.}(2020)\citenamefont {Khadka},
  \citenamefont {Thapaliya}, \citenamefont {Parra}, \citenamefont {Han},
  \citenamefont {Wen}, \citenamefont {Need}, \citenamefont {Khanal},
  \citenamefont {Wang}, \citenamefont {Zang}, \citenamefont {Kikkawa},
  \citenamefont {Wu},\ and\ \citenamefont {Huang}}]{khadka2020}%
  \BibitemOpen
  \bibfield  {author} {\bibinfo {author} {\bibfnamefont {D.}~\bibnamefont
  {Khadka}}, \bibinfo {author} {\bibfnamefont {T.~R.}\ \bibnamefont
  {Thapaliya}}, \bibinfo {author} {\bibfnamefont {S.~H.}\ \bibnamefont
  {Parra}}, \bibinfo {author} {\bibfnamefont {X.}~\bibnamefont {Han}}, \bibinfo
  {author} {\bibfnamefont {J.}~\bibnamefont {Wen}}, \bibinfo {author}
  {\bibfnamefont {R.~F.}\ \bibnamefont {Need}}, \bibinfo {author}
  {\bibfnamefont {P.}~\bibnamefont {Khanal}}, \bibinfo {author} {\bibfnamefont
  {W.}~\bibnamefont {Wang}}, \bibinfo {author} {\bibfnamefont {J.}~\bibnamefont
  {Zang}}, \bibinfo {author} {\bibfnamefont {J.~M.}\ \bibnamefont {Kikkawa}},
  \bibinfo {author} {\bibfnamefont {L.}~\bibnamefont {Wu}},\ and\ \bibinfo
  {author} {\bibfnamefont {S.~X.}\ \bibnamefont {Huang}},\ }\bibfield  {title}
  {\bibinfo {title} {Kondo physics in antiferromagnetic weyl semimetal
  {Mn}\textsubscript{3+x}{Sn}\textsubscript{1-x} films},\ }\href
  {https://doi.org/10.1126/sciadv.abc1977} {\bibfield  {journal} {\bibinfo
  {journal} {Science Advances}\ }\textbf {\bibinfo {volume} {6}},\ \bibinfo
  {pages} {eabc1977} (\bibinfo {year} {2020})}\BibitemShut {NoStop}%
\bibitem [{\citenamefont {Tsvelick}\ and\ \citenamefont
  {Wiegmann}(1983)}]{tsvelick1983}%
  \BibitemOpen
  \bibfield  {author} {\bibinfo {author} {\bibfnamefont {A.}~\bibnamefont
  {Tsvelick}}\ and\ \bibinfo {author} {\bibfnamefont {P.}~\bibnamefont
  {Wiegmann}},\ }\bibfield  {title} {\bibinfo {title} {Exact results in the
  theory of magnetic alloys},\ }\href
  {https://doi.org/10.1080/00018738300101581} {\bibfield  {journal} {\bibinfo
  {journal} {Advances in Physics}\ }\textbf {\bibinfo {volume} {32}},\ \bibinfo
  {pages} {453} (\bibinfo {year} {1983})}\BibitemShut {NoStop}%
\bibitem [{\citenamefont {Coleman}(1984)}]{coleman1984}%
  \BibitemOpen
  \bibfield  {author} {\bibinfo {author} {\bibfnamefont {P.}~\bibnamefont
  {Coleman}},\ }\bibfield  {title} {\bibinfo {title} {New approach to the
  mixed-valence problem},\ }\href {https://doi.org/10.1103/PhysRevB.29.3035}
  {\bibfield  {journal} {\bibinfo  {journal} {Phys. Rev. B}\ }\textbf {\bibinfo
  {volume} {29}},\ \bibinfo {pages} {3035} (\bibinfo {year}
  {1984})}\BibitemShut {NoStop}%
\bibitem [{\citenamefont {Bickers}\ \emph {et~al.}(1987)\citenamefont
  {Bickers}, \citenamefont {Cox},\ and\ \citenamefont {Wilkins}}]{bickers1987}%
  \BibitemOpen
  \bibfield  {author} {\bibinfo {author} {\bibfnamefont {N.~E.}\ \bibnamefont
  {Bickers}}, \bibinfo {author} {\bibfnamefont {D.~L.}\ \bibnamefont {Cox}},\
  and\ \bibinfo {author} {\bibfnamefont {J.~W.}\ \bibnamefont {Wilkins}},\
  }\bibfield  {title} {\bibinfo {title} {Self-consistent large-{N} expansion
  for normal-state properties of dilute magnetic alloys},\ }\href
  {https://doi.org/10.1103/PhysRevB.36.2036} {\bibfield  {journal} {\bibinfo
  {journal} {Phys. Rev. B}\ }\textbf {\bibinfo {volume} {36}},\ \bibinfo
  {pages} {2036} (\bibinfo {year} {1987})}\BibitemShut {NoStop}%
\bibitem [{\citenamefont {Otte}\ \emph {et~al.}(2008)\citenamefont {Otte},
  \citenamefont {Ternes}, \citenamefont {von Bergmann}, \citenamefont {Loth},
  \citenamefont {Brune}, \citenamefont {Lutz}, \citenamefont {Hirjibehedin},\
  and\ \citenamefont {Heinrich}}]{otte2008}%
  \BibitemOpen
  \bibfield  {author} {\bibinfo {author} {\bibfnamefont {A.~F.}\ \bibnamefont
  {Otte}}, \bibinfo {author} {\bibfnamefont {M.}~\bibnamefont {Ternes}},
  \bibinfo {author} {\bibfnamefont {K.}~\bibnamefont {von Bergmann}}, \bibinfo
  {author} {\bibfnamefont {S.}~\bibnamefont {Loth}}, \bibinfo {author}
  {\bibfnamefont {H.}~\bibnamefont {Brune}}, \bibinfo {author} {\bibfnamefont
  {C.~P.}\ \bibnamefont {Lutz}}, \bibinfo {author} {\bibfnamefont {C.~F.}\
  \bibnamefont {Hirjibehedin}},\ and\ \bibinfo {author} {\bibfnamefont {A.~J.}\
  \bibnamefont {Heinrich}},\ }\bibfield  {title} {\bibinfo {title} {The role of
  magnetic anisotropy in the {Kondo} effect},\ }\href
  {https://doi.org/10.1038/nphys1072} {\bibfield  {journal} {\bibinfo
  {journal} {Nature Physics}\ }\textbf {\bibinfo {volume} {4}},\ \bibinfo
  {pages} {847} (\bibinfo {year} {2008})}\BibitemShut {NoStop}%
\bibitem [{\citenamefont {Žitko}\ \emph {et~al.}(2009)\citenamefont {Žitko},
  \citenamefont {Peters},\ and\ \citenamefont {Pruschke}}]{zitko2009}%
  \BibitemOpen
  \bibfield  {author} {\bibinfo {author} {\bibfnamefont {R.}~\bibnamefont
  {Žitko}}, \bibinfo {author} {\bibfnamefont {R.}~\bibnamefont {Peters}},\
  and\ \bibinfo {author} {\bibfnamefont {T.}~\bibnamefont {Pruschke}},\
  }\bibfield  {title} {\bibinfo {title} {Splitting of the kondo resonance in
  anisotropic magnetic impurities on surfaces},\ }\href
  {https://doi.org/10.1088/1367-2630/11/5/053003} {\bibfield  {journal}
  {\bibinfo  {journal} {New Journal of Physics}\ }\textbf {\bibinfo {volume}
  {11}},\ \bibinfo {pages} {053003} (\bibinfo {year} {2009})}\BibitemShut
  {NoStop}%
\bibitem [{\citenamefont {Schlottmann}(1983)}]{schlottmann1983}%
  \BibitemOpen
  \bibfield  {author} {\bibinfo {author} {\bibfnamefont {P.}~\bibnamefont
  {Schlottmann}},\ }\bibfield  {title} {\bibinfo {title} {{Bethe-Ansatz}
  solution of the ground-state of the {SU} (2j+1) {Kondo (Coqblin-Schrieffer)}
  model: Magnetization, magnetoresistance and universality},\ }\href
  {https://doi.org/https://doi.org/10.1007/BF01307678} {\bibfield  {journal}
  {\bibinfo  {journal} {Zeitschrift f{\"u}r Physik B Condensed Matter}\
  }\textbf {\bibinfo {volume} {51}},\ \bibinfo {pages} {223} (\bibinfo {year}
  {1983})}\BibitemShut {NoStop}%
\bibitem [{\citenamefont {Coqblin}\ and\ \citenamefont
  {Schrieffer}(1969)}]{coqblin1969}%
  \BibitemOpen
  \bibfield  {author} {\bibinfo {author} {\bibfnamefont {B.}~\bibnamefont
  {Coqblin}}\ and\ \bibinfo {author} {\bibfnamefont {J.~R.}\ \bibnamefont
  {Schrieffer}},\ }\bibfield  {title} {\bibinfo {title} {Exchange interaction
  in alloys with cerium impurities},\ }\href
  {https://doi.org/10.1103/PhysRev.185.847} {\bibfield  {journal} {\bibinfo
  {journal} {Phys. Rev.}\ }\textbf {\bibinfo {volume} {185}},\ \bibinfo {pages}
  {847} (\bibinfo {year} {1969})}\BibitemShut {NoStop}%
\bibitem [{\citenamefont {Kondo}(2005)}]{kondo2005}%
  \BibitemOpen
  \bibfield  {author} {\bibinfo {author} {\bibfnamefont {J.}~\bibnamefont
  {Kondo}},\ }\bibfield  {title} {\bibinfo {title} {Sticking to my bush},\
  }\href {https://doi.org/10.1143/JPSJ.74.1} {\bibfield  {journal} {\bibinfo
  {journal} {Journal of the Physical Society of Japan}\ }\textbf {\bibinfo
  {volume} {74}},\ \bibinfo {pages} {1} (\bibinfo {year} {2005})}\BibitemShut
  {NoStop}%
\bibitem [{\citenamefont {Engelsman}\ \emph {et~al.}(1973)\citenamefont
  {Engelsman}, \citenamefont {Wiegers}, \citenamefont {Jellinek},\ and\
  \citenamefont {Van~Laar}}]{engelsman1973}%
  \BibitemOpen
  \bibfield  {author} {\bibinfo {author} {\bibfnamefont {F.~M.~R.}\
  \bibnamefont {Engelsman}}, \bibinfo {author} {\bibfnamefont {G.~A.}\
  \bibnamefont {Wiegers}}, \bibinfo {author} {\bibfnamefont {F.}~\bibnamefont
  {Jellinek}},\ and\ \bibinfo {author} {\bibfnamefont {B.}~\bibnamefont
  {Van~Laar}},\ }\bibfield  {title} {\bibinfo {title} {Crystal structures and
  magnetic structures of some metal(i) chromium(iii) sulfides and selenides},\
  }\href {https://doi.org/https://doi.org/10.1016/S0022-4596(73)80018-0}
  {\bibfield  {journal} {\bibinfo  {journal} {Journal of Solid State
  Chemistry}\ }\textbf {\bibinfo {volume} {6}},\ \bibinfo {pages} {574}
  (\bibinfo {year} {1973})}\BibitemShut {NoStop}%
\bibitem [{\citenamefont {Langari}\ and\ \citenamefont
  {Thalmeier}(2006)}]{Thalmeier2006}%
  \BibitemOpen
  \bibfield  {author} {\bibinfo {author} {\bibfnamefont {A.}~\bibnamefont
  {Langari}}\ and\ \bibinfo {author} {\bibfnamefont {P.}~\bibnamefont
  {Thalmeier}},\ }\bibfield  {title} {\bibinfo {title} {Antiferromagnetic and
  spin-gap phases of the anisotropic kondo necklace model},\ }\href
  {https://doi.org/10.1103/PhysRevB.74.024431} {\bibfield  {journal} {\bibinfo
  {journal} {Phys. Rev. B}\ }\textbf {\bibinfo {volume} {74}},\ \bibinfo
  {pages} {024431} (\bibinfo {year} {2006})}\BibitemShut {NoStop}%
\bibitem [{\citenamefont {Fitzgerald}\ \emph {et~al.}(1968)\citenamefont
  {Fitzgerald}, \citenamefont {Keil},\ and\ \citenamefont
  {Heinrich}}]{heinrich1968}%
  \BibitemOpen
  \bibfield  {author} {\bibinfo {author} {\bibfnamefont {R.}~\bibnamefont
  {Fitzgerald}}, \bibinfo {author} {\bibfnamefont {K.}~\bibnamefont {Keil}},\
  and\ \bibinfo {author} {\bibfnamefont {K.~F.~J.}\ \bibnamefont {Heinrich}},\
  }\bibfield  {title} {\bibinfo {title} {Solid-state energy-dispersion
  spectrometer for electron-microprobe x-ray analysis},\ }\href
  {https://doi.org/10.1126/science.159.3814.528} {\bibfield  {journal}
  {\bibinfo  {journal} {Science}\ }\textbf {\bibinfo {volume} {159}},\ \bibinfo
  {pages} {528} (\bibinfo {year} {1968})}\BibitemShut {NoStop}%
\end{thebibliography}
%

\section{Acknowledgements}

We thank A. P. Mackenzie, M. M. Piva, P. Thalmeier and U. Burkhardt for useful discussions, and S. Seifert for experimental support. J. Guimarães, D. Fartab and M. Moravec acknowledge support from the International Max Planck Research School for Chemistry and Physics of Quantum Materials (IMPRS-CPQM). We are grateful to the Max Planck Society for financial support.

\section{Author contributions}

J.G. and H.Z. conceived the research project. J.G., D.F. and H.Z. collaborated in the device fabrication. M.S. grew the single crystals. M.B. performed the magnetic susceptibility measurements. J.G. and H.Z. performed the electrical transport measurements, which were analysed by J.G.. J.G., M.M., B.S. and H.Z. were involved in many discussions to come up with the physical model that supported the observed data. H.Z. led the project and the paper was written by J.G., H.Z. and B.S., with input and comments by M.M. and M.B..

\section{Competing interests}

The authors declare no competing interests.

\end{document}